# Prior-information based super-resolution optical metrology of 2D nanoscale objects

Jin-Kyu So[1], Eng Aik Chan[1], Carolina Rendón-Barraza[1], Giorgio Adamo[2, 1] *, and Nikolay I. Zheludev[2,3]

[1]*Centre for Disruptive Photonic Technologies, SPMS, TPI, Nanyang Technological University, Singapore 637371.*

[2]*Optoelectronics Research Centre & Centre for Photonic Metamaterials, University of Southampton, SO17 1BJ, UK.*

*3. Hagler Institute for Advanced Study, Texas A&M University, College Station, Texas, 77843, USA*

**Correspondence to: g.adamo@soton.ac.uk.*

**Abstract**

Previous work has shown that optical metrology of one-dimensional objects, such as slit width, can achieve improved accuracy by using prior information from similar objects to train the metrology estimator. Here, we demonstrate single-shot optical metrology of nanoscale elliptical particles by analysing their diffraction patterns to retrieve length, width and in-plane orientation using a neural-network estimator trained on prior information from nano-ellipses with varied dimensions and orientations. Fisher-information flow analysis was used to optimise the physical parameters of the metrology apparatus and maximise measurement accuracy. Using a 633 nm laser, we measure the dimensions of elliptical particles with accuracy down to λ/128, corresponding to 4.9 nm, and recover their orientation with 5° accuracy. Our results demonstrate the practicality of optical, deep-super-resolution, single-shot, multiparameter measurements of two-dimensional subwavelength objects, with potential relevance to microbiology and nanotechnology applications.

## Introduction

When electromagnetic waves interact with an object, the scattered field[1] contains information about the object, enabling imaging[2–4], metrology[5,6], and broader characterization[7,8]. Estimating an object's physical parameters from the scattered light constitutes an ill-posed inverse scattering problem, solved by approximations and assumptions[9], whose accuracy depends critically on the amount of information that can be captured[10]. A neural network used as estimator can also solve the inverse scattering problems if trained on diffraction patterns from similar objects. This approach delivers deeply subwavelength accuracy in retrieving physical parameters or positions of nanoparticles[11–14], one-dimensional slits[15–18] and two-dimensional[19] nanoscale objects.

In this work, we develop a metrology method capable of retrieving multiple parameters of complex nanoscale objects. We demonstrate deeply subwavelength optical metrology of the long axis, short axis and azimuthal orientation of two-dimensional ellipses of subwavelength dimensions, a task which is impossible with conventional diffraction limited microscopes. Moreover, we reveal that the pattern of the Fisher information flow[1,20] depends on the measured parameter of the target and that there exists an optimal distance between the object and the imaging plane that allows to achieve the best estimation accuracy. We use the Fisher information analysis as a guide to design a single experimental setup configuration that allows to estimate all parameters with high accuracy. This experimental demonstration of multiparameter deep subwavelength metrology finds its importance in the widespread presence elliptical geometries have in both biology (e.g. viruses and cells), and nanotechnology (e.g. nanoparticles).

## Methodology of the optical scattering metrology

Optical scattering metrology retrieves the dimensions and orientation of a subwavelength object by analyzing the intensity distribution of its coherent light diffraction pattern. Although the task of retrieving parameters of the object from its intensity diffraction pattern constitutes an inherently ill-posed inverse problem, we address it by restricting the parameter space and using a deep learning–based metrology estimator - a convolutional neural network - trained on a large dataset of diffraction patterns generated from similar objects spanning a range of subwavelength dimensions and orientations. Information is acquired using an imaging camera that captures a partial subset of the scattered light field. Measurements are conducted within an optical microscope, where the metrology target is illuminated through a lens. The resulting diffraction pattern is then imaged and recorded at a distance $H$ up to a hundred wavelengths from the target, Fig.1b. We used a microscope with overall magnification of ×156 and a diode laser operating at $\lambda = 633$ nm.

Our metrology targets are gold ellipses of different sizes, aspect ratios and orientations manufactured by electron-beam lithography on glass substrates coated with conductive transparent indium tin oxide, Fig 1a. On a single wafer we fabricated 2000 ellipses with width $W$ ranging from $0.1\lambda$ to $0.9\lambda$, length $L$ ranging from $0.1\lambda$ to $0.6\lambda$ and orientation $\alpha$ ranging from 0° to 180°. Parameters of the ellipses were measured in the scanning electron microscope to create ground truth values for assortment of the optical scattering metrology. For each of the ellipses we recorded their diffraction patterns at the distances $H$ from the sample plane ranging from 0 λ to 100 λ. About 90% of the diffraction pattern images were used to train the neural network estimator. The trained neural network was then used to retrieve width, length and orientation of the unseen ellipses from their diffraction patterns. The estimator used to retrieve $L$, $W$, and $\alpha$ from the intensity diffraction patterns was ResNet-34[21], a 34 layers convolutional neural network, widely used for image recognition.

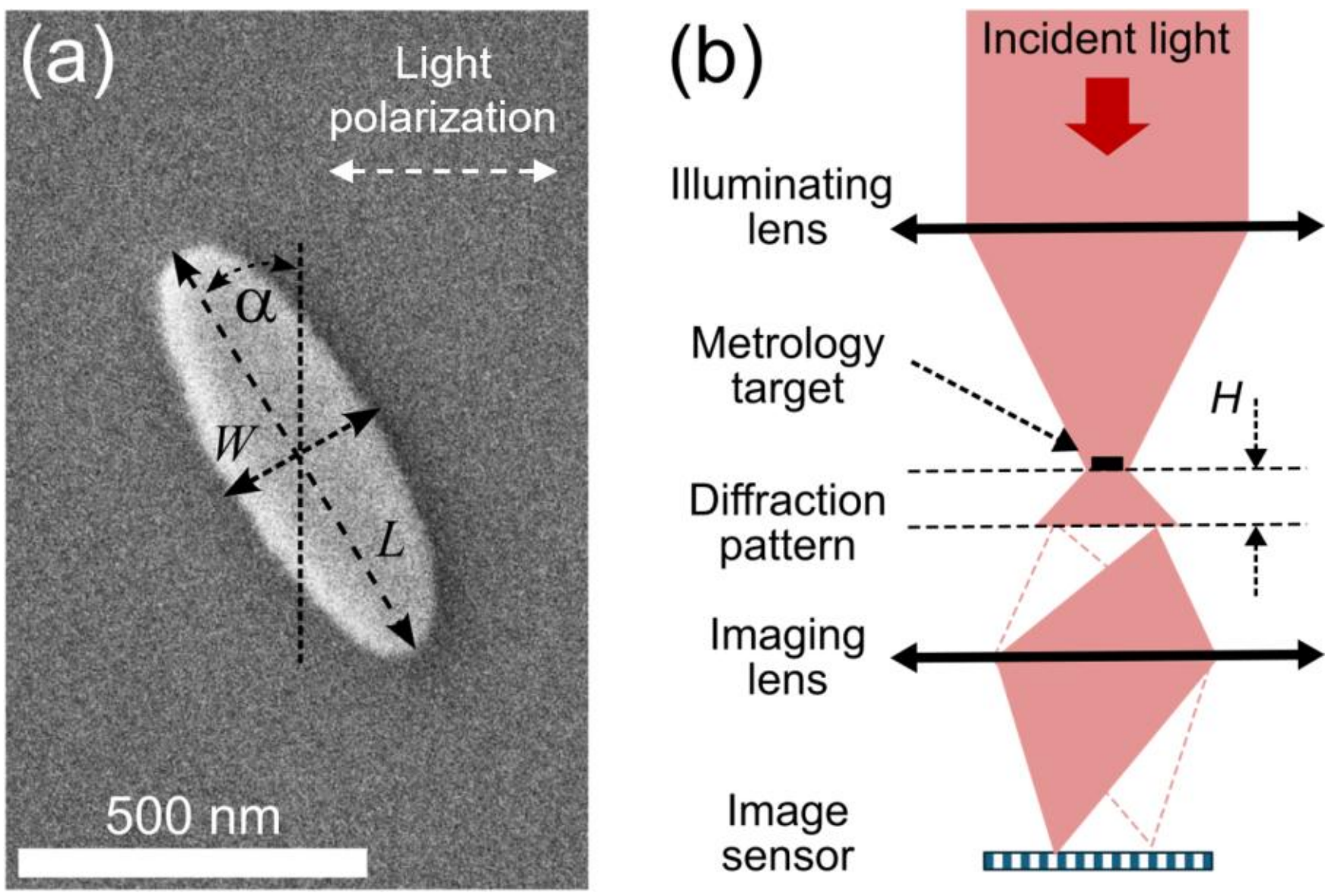

**Figure 1. Optical scattering metrology of nanoparticles.** (a) SEM image of a representative elliptical nanoparticle with its geometrical parameters, $L$, $W$, and $\alpha$ labeled. (b) A sketch of the optical scattering metrology apparatus.

## Fisher information guided experimental configuration

As our experiments target to estimate several parameters using a single configuration of the optical setup, we adopt the concept of Fisher information flux[1,20] to understand how the information related to a particular parameter of the ellipses flows toward the imaging camera. Full-wave electromagnetic simulations (Lumerical FDTD) were used to calculate energy and Fisher information fluxes for an ellipse of size $L$ = 600 nm and $W$ = 120 nm, illuminated at normal incidence by a plane-wave. Figure 2 shows the radiation patterns of the energy flux $\vec{S}$ and the Fisher information flux[22] $\overrightarrow{S^{FI}}$ for a nanoscale ellipse lying in the $x$-$y$ plane, under $y$-polarized incident light propagating along the $z$-axis. The top, middle, and bottom rows correspond to ellipse orientation angles of $\alpha = 0°$, 45°, and 90°, respectively. The energy flux patterns are shown in Figs. 2a, 2e, and 2i, while the Fisher information flux patterns for the three ellipse parameters are shown in Figs. 2b-d, 2f-2h, and 2j-2l.

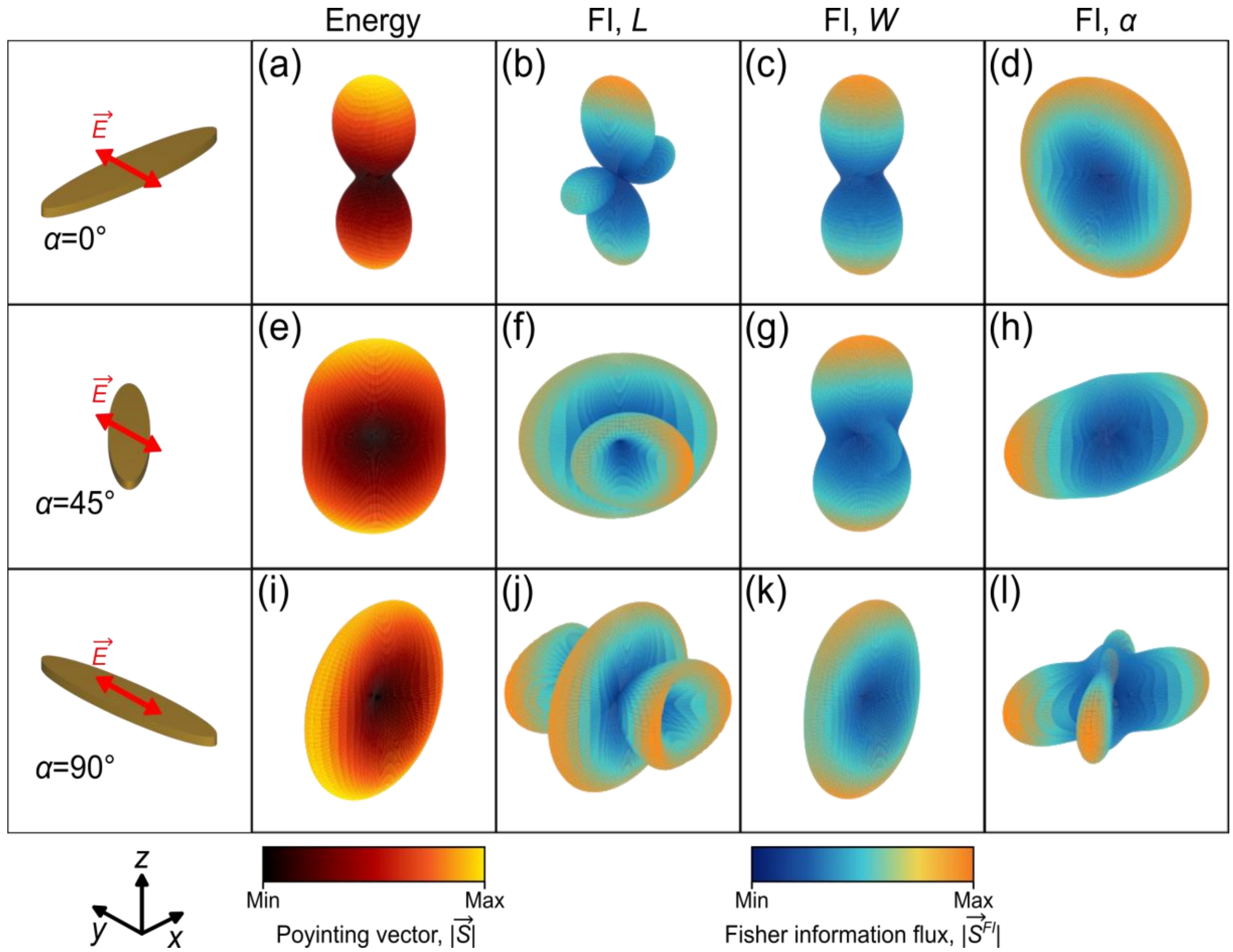

**Figure 2. Patterns for energy and Fisher information flux scattered from a nanoellipse**. The particle is placed on the *x*-*y* plane and illuminated by a *y*-polarized plane-wave propagating along *z*, with ellipse orientation angles α = 0° (top row), 45° (middle row), and 90° (bottom row), measured from *x*-axis.

From the radiation patterns, it stems clearly that Fisher information radiation flux for *L*, *W*, and *α* do not overlap with the energy flux, particularly in the case of the ellipses' orientation. Given the directionality of the information flux for the three parameters, placing the imaging detector along the light propagation direction (*z* axis) appears to be the optimal choice for their concurrent estimation using a fixed experimental arrangement. The shapes of the radiation patterns and the presence of side lobes suggest that highest accuracy shall be expected in the estimation of *W* while the lowest in the estimation of *α*.

### Experimental estimation accuracy for *W*, *L*, *α*

We allocated 80%, 10%, and 10% of the 2,000 experimental diffraction patterns to the training, validation, and test of the neural network, respectively. We trained and evaluated 25 independent neural networks, using randomized versions of the datasets. The estimated values for the *L*, *W* and *α* were obtained by averaging the predictions across all networks. Randomizing the geometrical parameters prevented the neural networks from learning correlations arising from background intensity patterns.

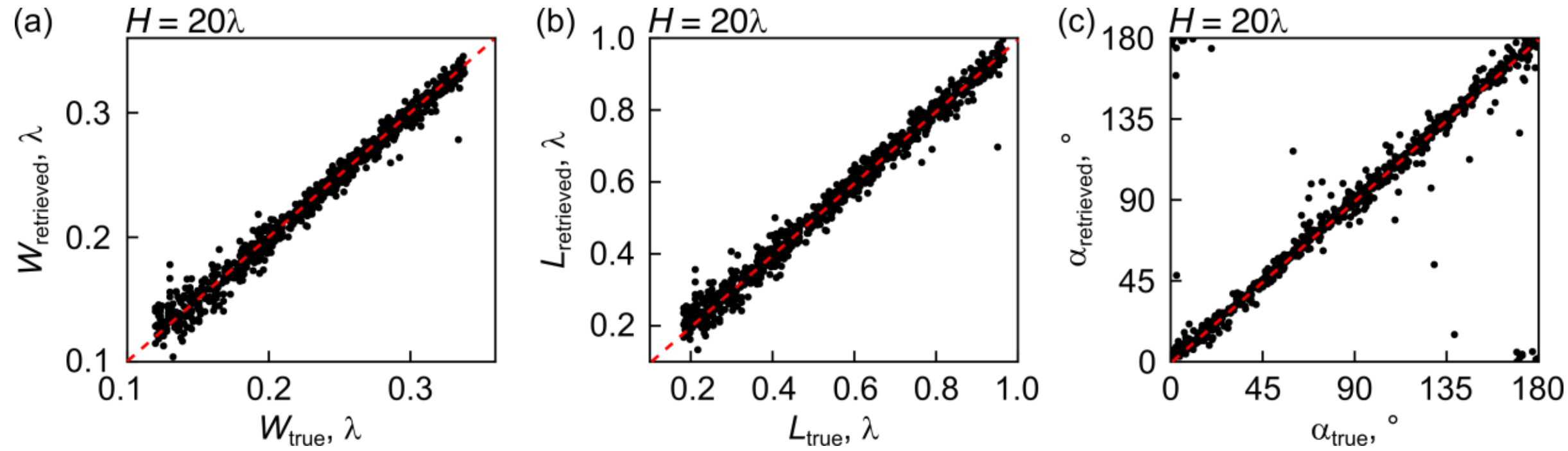

**Figure 3. Experimental estimation accuracy of ellipses parameters.** The retrieved values of the ellipses (a) width $W_{retrieved}$, (b) $L_{retrieved}$, and (c) $\alpha_{retrieved}$—obtained from intensity diffraction patterns at $H = 20$ λ of 200 nanoparticles under Gaussian beam illumination (FWHM=1.1 λ)—are plotted against the ground truths for same parameters measured with SEM, $W_{true}$, $L_{true}$, $\alpha_{true}$.

The estimation accuracies for the three parameters, obtained from test experiments on 200 randomly selected ellipses, are presented in Fig. 3. The plots show the optically retrieved values of ellipses length, width and orientation, measured at a distance $H = 20\ \lambda$ from the sample, as a function of the corresponding ground truth values, obtained by SEM measurements (see also supplementary Figure S1). The dashed lines indicate perfect agreement, while deviation from the line quantifies the measurement accuracy. Figure 3(a) shows the measurements results for the length, Fig. 3(b) for the width, and Fig. 3(c) for the rotation angle. The corresponding root mean square values are $\delta W_{RMS} = \lambda/128$ (4.9 nm), $\delta L_{RMS} = \lambda/38$ (16.7 nm), and $\delta\alpha_{RMS} = 5.1°$. Consistent with the analysis of the Fisher information flux[1,21], we observe the highest accuracy in retrieving the ellipses width, $W$, a lower accuracy for the length, $L$, and the lowest accuracy for the rotation angle, $\alpha$.

**Dependence of measurement accuracy on the sampling distance**

The neural network estimates the ellipse parameters from information encoded in the diffraction intensity pattern. As light propagates away from the target, the pattern diverges and evolves. Since the detector has finite size and pixel resolution, the captured portion of the pattern, signal-to-noise ratio and the resolved detail of its features both depend on the sampling distance $H$. Together, these effects produce a nonmonotonic dependence of estimation accuracy on $H$.

To assess this dependence, we repeated the metrology measurements of the elliptical target over a range of sampling distances from 0 to 100λ. Figure 4a shows a 69λ x 69λ region of the diffraction patterns captured at several sampling distances by an image sensor over 1047 × 1047 pixels and a magnifying imaging lens, as illustrated in Figure 1b. The root mean square values, $\delta W_{RMS}$, $\delta L_{RMS}$, and $\delta\alpha_{RMS}$, initially decrease as the imaging distance increases, reach a minimum, and then increase again at larger distances with optimal accuracy reached at $H = 15$-$20$ λ (Fig. 4b, c).

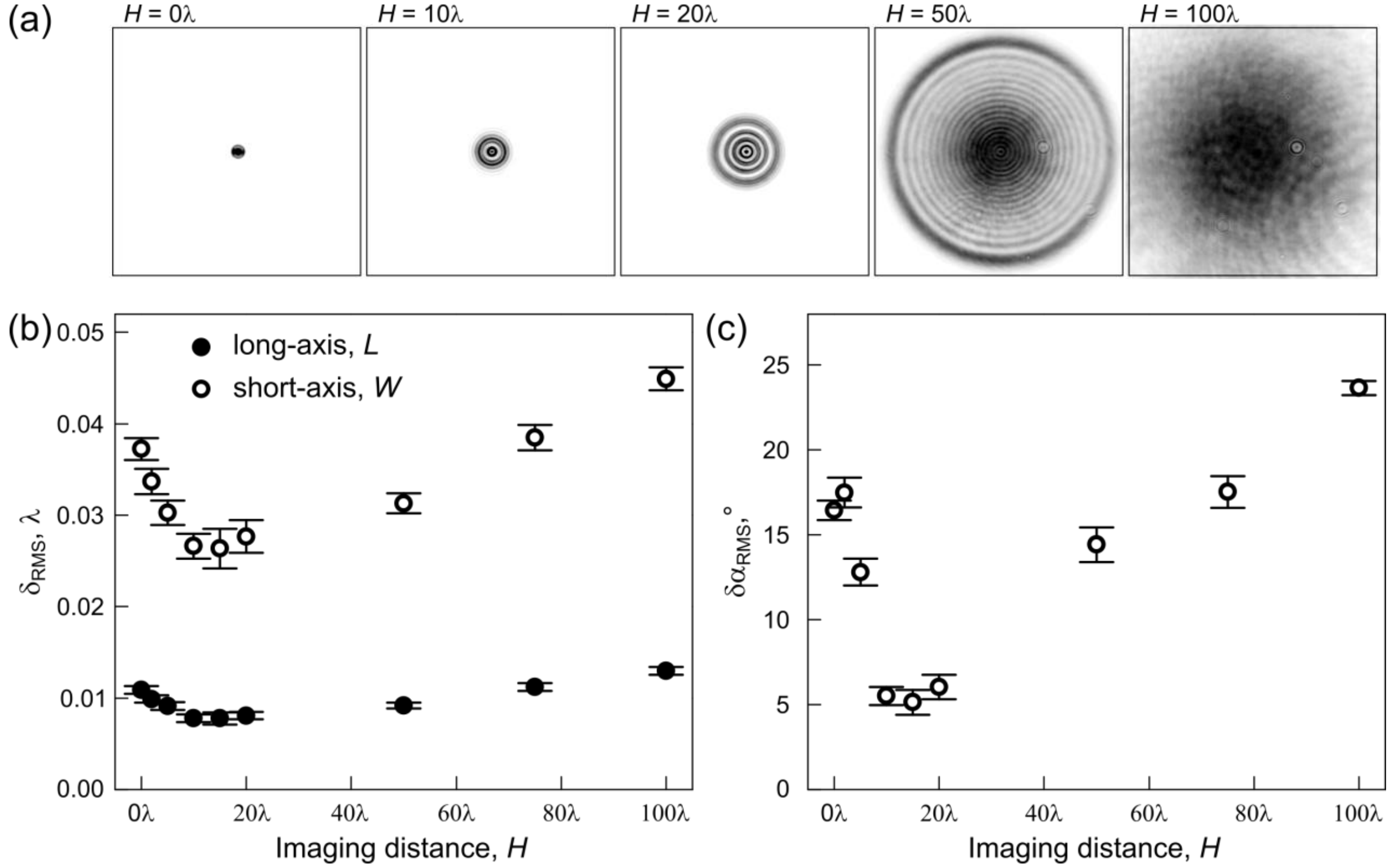

**Figure 4. Dependence of measurement accuracy on imaging distance.** (a) Intensity diffraction patterns taken at imaging distances, $H$ = 0, 10, 20, 50, 100 λ from the sample, with a FOV = 69.1 λ × 69.1 λ. (b-c) Measurement accuracy of the ellipses (b) width, $\delta W_{RMS}$, length, $\delta L_{RMS}$, and (c) orientation, $\delta\alpha_{RMS}$, as function of imaging distance, $H$.

## Conclusion

In summary, we developed a deeply subwavelength single-shot non-invasive optical metrology approach for measuring multiple independent dimension and orientation parameters of a two-dimensional nanoscale object. The method uses scattering metrology optimized by analyzing the Fisher information flow generated by the target and is enabled by training an estimator on intensity diffraction patterns from similar objects. It achieves measurement accuracy better than one percent of the optical wavelength. These results demonstrate the practicality of optical metrology for multiparameter measurements of subwavelength objects, with potential applications in microbiology and nanotechnology.

## Acknowledgments

This work was supported by the Singapore National Research Foundation (Grant NRF-CRP23-2019-0006) and the UK Engineering and Physical Sciences Research Council (grants EP/T02643X/1 and EP/Z53285X/1)

## Author contribution

J.-K. S., G. A. and N. I. Z. conceptualized the project. J.-K. S. assembled the apparatus, fabricated the samples and performed the measurements. C. E. A. developed the neural network for the image analysis and reconstruction. C. R.-B. developed the automated image acquisition program. J.-K. S. performed the simulation of Fisher information flow. C. E. A. performed the Fisher information analysis of the experimental data with inputs from J.-K. S. All authors contributed to data analysis and interpretation. J.-K. S. prepared the initial manuscript draft, which was jointly revised by all authors. N. I. Z. and G.A. supervised the project.

## Competing interests

The authors declare no competing interests.

## Data availability

The data that support the findings of this study are openly available from the NTU research data repository DR-NTU (Data).

## References

1. Hüpfl, J. *et al.* Continuity equation for the flow of Fisher information in wave scattering. *Nature Physics* **20**, 1294–1299 (2024).
2. Jiang, Y. *et al.* Electron ptychography of 2D materials to deep sub-ångström resolution. *Nature* **559**, 343–349 (2018).
3. Katz, O., Heidmann, P., Fink, M. & Gigan, S. Non-invasive single-shot imaging through scattering layers and around corners via speckle correlations. *Nature Photon* **8**, 784–790 (2014).
4. Ben-Yehuda, A. *et al.* High-resolution computed tomography with scattered X-ray radiation and a single pixel detector. *Commun Eng* **3**, 39 (2024).
5. Orji, N. G. *et al.* Metrology for the next generation of semiconductor devices. *Nat Electron* **1**, 532–547 (2018).
6. Piliarik, M. & Sandoghdar, V. Direct optical sensing of single unlabelled proteins and super-resolution imaging of their binding sites. *Nat Commun* **5**, 4495 (2014).
7. Sarenac, D. *et al.* Small-angle scattering interferometry with neutron orbital angular momentum states. *Nat Commun* **15**, 10785 (2024).
8. Hollamby, M., Hanayama, H. & Yagai, S. Using in-situ small-angle scattering to reveal the structure and dynamics of supramolecular polymers. *Nat Commun* **16**, 9316 (2025).
9. Lee, M., Hugonnet, H. & Park, Y. Inverse problem solver for multiple light scattering using modified Born series. *Optica, OPTICA* **9**, 177–182 (2022).
10. Kay, S. M. *Fundamentals of Statistical Signal Processing: Estimation Theory*. (Prentice Hall, 1993).
11. Wang, Y. *et al.* 3D positional metrology of a virus-like nanoparticle with topologically structured light. *Applied Physics Letters* **124**, (2024).

12. Wang, B. *et al.* Retrieving positions of closely packed subwavelength nanoparticles from their diffraction patterns. *Applied Physics Letters* **124**, (2024).

13. Hu, X. *et al.* Deep-learning-augmented microscopy for super-resolution imaging of nanoparticles. *Optics Express* **32**, 879–890 (2023).

14. Chan, E. A. *et al.* Counting and mapping of subwavelength nanoparticles from a single shot scattering pattern. *Nanophotonics* **12**, 2807–2812 (2023).

15. Pu, T., Ou, J.-Y., Papasimakis, N. & Zheludev, N. I. Label-free deeply subwavelength optical microscopy. *Applied Physics Letters* **116**, (2020).

16. Rendón-Barraza, C. *et al.* Deeply sub-wavelength non-contact optical metrology of sub-wavelength objects. *APL Photonics* **6**, (2021).

17. Pu, T. *et al.* Unlabeled Far-Field Deeply Subwavelength Topological Microscopy (DSTM). *Advanced Science* **8**, 2002886 (2021).

18. Liu, T. *et al.* Picophotonic localization metrology beyond thermal fluctuations. *Nat. Mater.* **22**, 844–847 (2023).

19. Wang, Y. *et al.* 2D Super-Resolution Metrology Based on Superoscillatory Light. *Advanced Science* 2404607 (2024).

20. Weimar, M. *et al.* Controlling the Flow of Information in Optical Metrology. Preprint at https://doi.org/10.48550/arXiv.2508.13640 (2025).

21. He, K., Zhang, X., Ren, S. & Sun, J. Deep residual learning for image recognition. in 770–778 (2016).

22. Yang, J., Hugonin, J.-P. & Lalanne, P. Near-to-Far Field Transformations for Radiative and Guided Waves. *ACS Photonics* **3**, 395–402 (2016).

23. Loshchilov, I. & Hutter, F. Decoupled Weight Decay Regularization. in (2018).

## Methods

### Experimental optical setup and measurement procedure

Linearly polarized laser light, of wavelength $\lambda$ = 633 nm, illuminated the samples through a microscope objective lens (NA = 0.9), with Gaussian beam profile of FWHM = 1.1 $\lambda$. The resulting diffraction patterns were collected by another microscope objective lens (NA = 0.9) at various distances, $H$, away from the sample plane, and imaged on a sCMOS camera (Andor Neo). The overall magnification of the imaging system was ×156, corresponding to an effective pixel size of 41.7 nm on the sample plane. The exposure time was set so that the peak counts at each imaging distance were similar, while not saturating the image. The effective pixel size of the imaging system was 41.7 nm in the sample plane. The samples were oriented so that the linear polarization of the incident laser light was orthogonal to the long-axis of the nanoellipses when $\alpha = 0°$. For characterization purposes, the nanoparticles were arranged in arrays of 10 x 10, and the sample position was calibrated for each array using markers fabricated during the same EBL process.

### Artificial neural network

We utilized a 34-layer Residual Network (ResNet-34) pretrained with the ImageNet v1 dataset (IMAGENET1K_V1)[20] . The network was trained using the AdamW optimizer[23] with a learning rate of 0.0003 and weight decay set to 0.02, alongside default beta values of (0.9, 0.999). To optimize training efficiency, we employed a multi-step learning rate scheduler (MultistepLR) that decayed the learning rate by a factor of 0.8 every 20 epochs.

The input to the network comprises of 16-bit diffraction intensity images, with the output being the metrology parameters. Prior to feeding into the network, the input images were resized to 384 pixels by 384 pixels, ensuring preservation of diffraction features without loss in diffraction intensity resolution.

The diffraction images are pre-processed such that the intensity across different samples (for the same diffraction distance) are preserved and the maximum intensity of all samples are kept at 80% of the dynamic range (16 bits). The images are then saved as 16bits .tif. Then the network loads all the .tif images in trained set and calculate the mean ($\mu$) and standard deviation ($\sigma$) of pixels values across all training samples. During training as well as testing, the network will then subtract this calculated mean and divide by the calculated standard deviation before inputting into the neural network with the formula: $y = (x - \mu)/\sigma$, where $y$ is the pixels of the image input into the neural network and $x$ is the loaded images pixel.

Training optimization was performed using a mean absolute error loss function, which computes the absolute differences between the predicted and ground truth values of the parameters of interest. A batch size of 32 was chosen for training. The entire network was implemented using the PyTorch framework.

Training was conducted over 1,500 epochs, utilizing a single A100 GPU and spanning a total duration of 5 hours. For dataset partitioning, 80% was allocated for training, 10% for validation during training, and the remaining 10% for testing. To mitigate selection bias, each measurement was repeated 25 times, with randomized apportionment of the training, validation, and testing sets. Subsequently, expected values and standard deviation errors were computed from the 25 repetitions.

**Fisher information calculation**

To calculate the Fisher information, images at each $H$ were stacked as a function of the parameter, $\theta$. Then, Fisher information is calculated for each $H$ based on the formula:

$$F = \iint d\boldsymbol{A} \left(\frac{\partial \ln p}{\partial \theta}\right)^{\mathbf{2}} p$$

where $p$ is the probability density function of the indirect measurement (photon counts in each pixel of the diffraction pattern whose total counts are normalized to a single photon), $\theta$ is the parameter to be measured (orientation of the elliptical nanoparticle), $\boldsymbol{A}$ is the area of diffraction pattern. In practice, the derivative $\partial \ln p/\partial\theta$ was evaluated numerically using finite differences method across the parameter grid, and the integration over A was performed using numerical Romberg integration on the image pixel grids. Then, the average Fisher information was calculated with respect to $\theta$, $(\sum_\theta F)/N_\theta$, where $N_\theta$ is the number of $\theta$ used in the calculation**.** The average value of Fisher information, $F(\theta)$, was used to get the Cramer-Rao bound, $(1/F)^{1/2}$ at each imaging distance, $H$. Since the Fisher information calculated for a single photon diminishes with $\sqrt{n}$ where n is the number of photons, the final Fisher information was divided by $\sqrt{n}$ for each $H$.

**Rescaling of predictions and ground truth**

To expedite convergence during training, all predictions and ground truth values, $P$ except for angle retrieval, were normalized to a range of 0 to 1 using the formula:

$$P_{norm} = \frac{P - P_{min}}{P_{max} - P_{min}}$$

For angle retrieval, the angle α was represented using a 2-vector Euler's trigonometric representation:

$$P_{angle} = (cos(2\alpha), sin(2\alpha))$$

, which ensures the angle function remains continuous between the two extremes of angles when θ = 0° and 180°.

**Fabrication of nanoparticles**

A glass coverslip (2.4mm x 1.2mm) was cleaned with Acetone/IPA/DI water in ultrasonic cleaner. A 40-nm-thick ITO was sputtered on the clean glass coverslip at forward RF power of 24 W. Bottom resist (PMMA 950K, AR-P 672.045, 250nm) was spin coated and baked on a hotplate at 150 °C for 3 min. Top resist (CSAR 62, AR-P 6200.09, 200nm) was spin coated and baked on a hotplate at 150 °C for 1 min. Electron beam exposure (30 keV, 10.3 pA) was done with a point dose of 1500 µC. After the resist developed, a 50nm-thick gold film with a 2nm-thick Ti adhesion layer was deposited with a thermal evaporator and lift-off was done with a remover.

**Nanoparticle positioning under the optical microscope**

To ensure accurate positioning of each nanoparticle, a 10 x 10 array of circular particles of diameter = 1 µm with interspacing of 15 µm was prepared together with marker particles to correct the distortion in the writing field of the scanning electron microscope used for electron beam lithography. Starting from one of marker particles, images of particles were taken while moving the microscope stage based on nominal positions defined in the CAD design. Then, the particle center of each image was retrieved and new positions of each particle were obtained to compensate the offset between the retrieved center of the reference marker and each particle. This was repeated until the positioning accuracy is converged.

# Supporting Information

## Prior-information based super-resolution optical metrology of 2D nanoscale objects

Jin-Kyu So[1], Eng Aik Chan[1], Carolina Rendón-Barraza[1], Giorgio Adamo[2, 1] *,

and Nikolay I. Zheludev[2,3]

[1]Centre for Disruptive Photonic Technologies, SPMS, TPI, Nanyang Technological University, Singapore 637371.

[2]Optoelectronics Research Centre & Centre for Photonic Metamaterials, University of Southampton, SO17 1BJ, UK.

3. Hagler Institute for Advanced Study, Texas A&M University, College Station, Texas, 77843, USA

**Preparation of ground truth**

The ground truth provided for neural network training was prepared by first taking the SEM images of 100 nanoellipses fabricated in the same batch as the samples for the datasets. Based on the SEM images and measured lengths of 100 nanoellipses, a calibration was conducted between the actual and CAD-prescribed lengths as shown in Figure S1. The ground truth for the actual datasets was prepared based on this calibration.

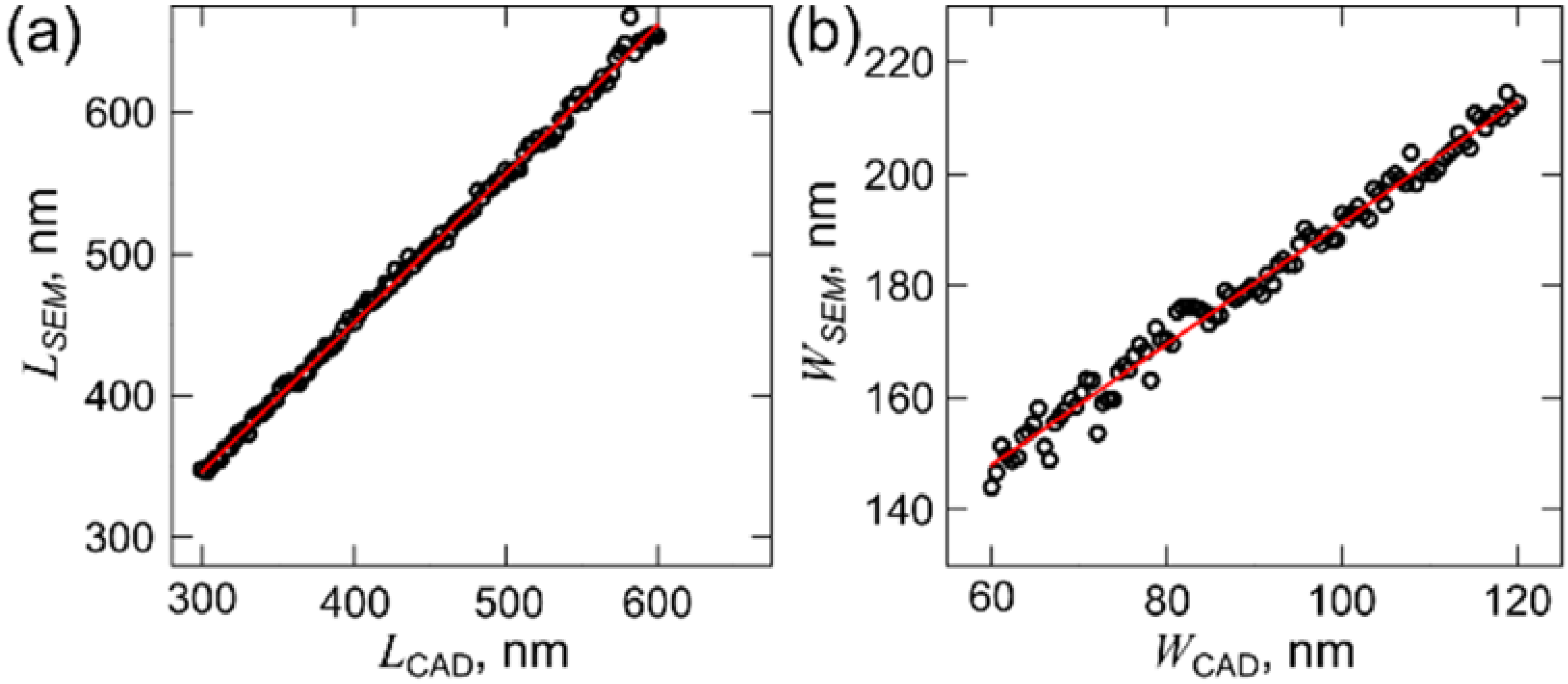

**Figure S1.** Comparison of SEM measured and design values of (a) long-axis and (b) short-axis of 100 elliptical nanoparticles for size calibration.